# Experimental demonstration of random walk by probability chaos using single photons

Makoto Naruse[1*], Martin Berthel[2], Hirokazu Hori[3], Aurélien Drezet[2], and Serge Huant[2]

[1] Department of Information Physics and Computing, Graduate School of Information Science and Technology, The University of Tokyo, 7-3-1 Hongo, Bunkyo-ku, Tokyo 113-8656, Japan

[2] Université Grenoble Alpes, CNRS, Institut Néel, 38000 Grenoble, France

[3] Interdisciplinary Graduate School, University of Yamanashi, Takeda, Kofu, Yamanashi 400-8510, Japan

*Email: makoto_naruse@ipc.i.u-tokyo.ac.jp

**Abstract**

In our former work (Sci. Rep. 4: 6039, 2014), we theoretically and numerically demonstrated that chaotic oscillation can be induced in a nanoscale system consisting of quantum dots between which energy transfer occurs via optical near-field interactions. Furthermore, in addition to the nanoscale implementation of oscillators, it is intriguing that the chaotic behavior is associated with probability derived via a density matrix formalism. Indeed, in our previous work (Sci. Rep. 6: 38634, 2016) we examined such oscillating probabilities via diffusivity analysis by constructing random walkers driven by chaotically driven bias. In this study, we experimentally implemented the concept of probability chaos using a single-photon source that was chaotically modulated by an external electro-optical modulator that directly yielded random walkers via single-photon observations after a polarization beam splitter. An evident signature was observed in the resulting ensemble average of the time-averaged mean square displacement. Although the experiment involved a scaled-up, proof-of-concept model of a genuine nanoscale oscillator, the experimental observations clearly validate the concept of oscillating probability, paving the way toward future ideal nanoscale systems.

Energy transfer based on optical near-field interactions has been intensively studied theoretically[1] and experimentally.[2,3] The optical excitation generated in the energy level of the smaller quantum dot, denoted by $S_1$ in Fig. 1(a), can be transferred to the upper energy level of the larger dot $L_2$ through optical near-field interactions.[4] The excitation on $L_2$ could then be shifted to the lower energy level $L_1$ via energy dissipation. Such near-field processes have been utilized in lighting and energy applications.[5] Additionally, fundamental studies toward computing and intelligent functions such as solution searching[6], decision making[7], and computing[8] are emerging to benefit from the novel optical mechanisms inherent in the subwavelength scale. In the context of functional systems, the generation of periodic signals is one of the most fundamental aspects of the functional systems observed in nature, such as heart beats, and artificially constructed devices, such as clock signals for microprocessors. Indeed, Naruse *et al.* theoretically demonstrated optical pulsation using energy transfer among quantum dots interacted via near-field interactions[9] by incorporating a delay mechanism as schematically indicated by $\Delta$ in Fig. 1(a).

Furthermore, combined with additional external delay, indicated by $\Delta_C$ in Fig. 1(a), chaotic oscillations and random number generations have been demonstrated.[10] The solid curves in Figs. 1(b,i), 1(b,ii), 1(b,iii), and 1(b,iv) show the time evolution of the population associated with the lower energy level of the larger quantum dot $L_2$ when the coupling strength $\alpha_C$ between the external delay system and the original energy transfer system is given by $30 \times 10^{-3}$, $180 \times 10^{-3}$, $182 \times 10^{-3}$, and $207 \times 10^{-3}$, respectively.[10] In Fig. 1(b), periodically oscillating evolution of the population in (i), chaotically evolving populations in (ii) and (iii), and a quasiperiodically evolving population in (iv) are evident. Let α$_C$ be a control parameter, and the red circular and the blue x marks in Fig. 1(c) represent local maxima and local minima of the populations, respectively. Bifurcations are clearly observed, which were adapted from Fig. 3 of our former work.[10]

In addition to the nanoscale implementation of such oscillatory behavior, it is intriguing that the observed time evolutions are associated with the elements of a density matrix, meaning that what is oscillating is probability. To examine the nature of such oscillatory probability, in the study in Ref. 11, we conducted a diffusivity analysis by constructing random walkers driven by dynamically changing bias supplied by the time evolution of the populations calculated from the quantum-dot systems involving near-field interactions and time delay. We examined the ensemble average of the time-averaged mean square displacement (ETMSD) of the random walkers, defined as

$$\text{ETMSD}(\Delta) = \left\langle \frac{1}{T-\Delta} \sum_{t=1}^{T-\Delta} \Delta x_s(t;\Delta)^2 \right\rangle_s, \quad (1)$$

where $\Delta x_s(t;\Delta) = x_s(t+\Delta) - x_s(t)$ means the displacement of the walker and $\langle \bullet \rangle_s$ denotes an ensemble average. One of the major results therein was that the ETMSD underwent large deviations from linear scaling in short time intervals in the case of chaotic oscillations, exhibiting a super-diffusion-like behavior. Such large diffusivity implies benefits in applications such as decision making; for example, the chaotic time series generated by a semiconductor laser provide efficient solutions to the multi-armed bandit problem.[12]

Based on these earlier works, we experimentally implemented the concept of dynamically changing probability, including chaotic changes, in this study by using a single-photon source modulated by an external electrooptic modulator. The single photon was then subjected to a polarization beam splitter (PBS) and detected by either of the two avalanche photodiodes (APDs) corresponding to horizontal or vertical polarization. As explained in detail below, the identity of the APD corresponds directly to the direction of the random walker. Consequently, the diffusivity is evaluated directly from the photon measurement sequences. Although the experiment was a scaled-up, proof-of-concept model of the genuine nanoscale oscillators, the experimental observations clearly validate the concept of oscillating probability, paving the way for future nanoscale systems.

The experimental setup is schematically illustrated in Fig. 2. A single photon is emitted from a single individual nitrogen-vacancy (NV) color center,[13] which has a broadband emission spectrum in the visible range (640–700 nm),[14–16] excited by a green laser (CNI Lasers, diode-pumped solid-state laser; Wavelength: 532 nm). After passing through a polarizer (denoted by Pol in Fig. 2) and a zero-order half-wave plate ($\lambda/2$), the emitted photon impinges on an electro-optical phase modulator (Thorlabs, EO-PM-NR-C1) driven by an external system through an amplifier. The modulated signal is subjected to the PBS, followed by single-photon detection by either $APD_H$ or $APD_V$ (Excelitas Technologies, SPCM-AQRH-16-FC), which correspond to horizontal and vertical polarization, respectively. The signal is then sent to a time-correlated single-photon-counting system (PicoQuant, PicoHarp 300).

Assume that there is no modulation by the modulator and that the polarization of the input single photon is 45° with respect to the horizontal axis; the photon is detected by $APD_H$ or $APD_V$ with 50:50 probability. Next, the single photons are modulated by a periodically oscillating signal, which is shown in Fig. 1(b,i), using the electro-optical phase modulator. In this study, the time scale of the nanoscale simulations in Fig. 1(b) was stretched by about a factor of $10^7$ in the driving signal in the experiment taking into account the experimentally available single-photon rate (which is on the order of $10^4$/s). The red and green traces in Fig. 3(a) show the number of photons per 2-ms-duration time bin detected by $APD_H$ and $APD_V$, respectively. The total duration of the measurement was about 12.7 s. Figure 3(b) is a magnified view of Fig. 3(a) corresponding to the initial 1 s. From these images, we can observe that the polarization of the incoming single photons is modulated by the periodic signals, while photons are detected simultaneously at either $APD_H$ or $APD_V$, especially when the periodical modulation signal crosses the zero point. That is, the probabilistic attributes of single photons are also observed.

Precisely, the photons were detected as schematically shown in Fig. 3(c), where single-photon detection events are depicted by red and green x marks, which correspond to detection by $APD_H$ and $APD_V$, respectively. In discussing the diffusivity induced by chaotically driven bias, we constructed a random walker directly from the single-photon detection events using the APDs. Specifically, the position of the random walker was updated by

$$x(t+1) = \begin{cases} x(t)+1 & \text{if photon is detected by } APD_H \\ x(t)-1 & \text{if photon is detected by } APD_V \end{cases}, \quad (2)$$

assuming the initial position of the random walker to be zero ($x(0) = 0$). Here, the time $t$ is updated when a single photon is detected by either of the detectors.

We prepared five kinds of modulation signal trains subjected to the optical phase modulator based on the numerical study concerning near-field-mediated energy transfer in Ref. 10:

[(1) Constant] Constant value: The polarization of single photons is maintained at about 45° with respect to the horizontal.

[(2) Periodic] A periodic signal specified by $\alpha_C = 30 \times 10^{-3}$, which is referred to as "30" hereafter, calculated in the near-field energy transfer calculations. (See Fig. 1(b,i).)

[(3) Quasiperiodic] A quasiperiodic signal specified by $\alpha_C = 207 \times 10^{-3}$ (called "207"). (See Fig. 1(b,iv).)

[(4) Chaos 1] A chaotic signal specified by $\alpha_C = 180 \times 10^{-3}$ (called "180"). (See Fig. 1(b,ii).)

[(5) Chaos 2] A chaotic signal specified by $\alpha_C = 182 \times 10^{-3}$ (called "182"). (See Fig. 1(b,iii).)

For each of the signal trains ((1)–(5)), the measurements were conducted 10 times. From each measurement $s$, a walker $x_s(t)$ was generated based on the rule given by Eq. (2), followed by calculating the time-averaged mean square displacement, which is the inner component of Eq. (2) given by

$$\text{TAMSD}_s(\Delta) = \frac{1}{T-\Delta} \sum_{t=1}^{T-\Delta} \Delta x_s(t;\Delta)^2, \tag{3}$$

whose ensemble average yields ETMSD: $\text{ETMSD}(\Delta) = \langle \text{TAMSD}_s(\Delta) \rangle_s$.

As is well known, when a random walker moves toward the plus and minus directions with probabilities $p$ and $q$, respectively, the mean displacement is given by $\langle x(t)\rangle = (p-q)t$, while the variance is given by $\langle (x(t)-\langle x(t)\rangle)^2\rangle = 4pqt$. Therefore, if the probability of single-photon detection by APD$_H$ and APD$_V$ is 50:50, the mean square displacement at time $t$ is equal to $t$ since $p = q = 1/2$, which is called normal diffusion. The black solid line in Fig. 4(a) depicts normal diffusion, meaning that the ETMSD is equal to the time difference $\tau$. The magenta marks in Fig. 4(a) show the ETMSD when the modulation is a constant value [(1) Constant], which almost perfectly agree with the normal diffusion. The red circular marks correspond to the case with periodically oscillating modulation [(2) Periodic], where the ETMSD is clearly modulated in a sinusoidal manner. The green, blue, and cyan marks correspond to the modulation by [(3) Quasiperiodic], [(4) Chaos 1], and [(5) Chaos 2], respectively, which quickly deviate from the normal diffusion. Especially with chaotic modulation, the ETMSD exceeds 250 around $\tau = 9$ s, which is larger than the normal diffusion by a factor of about 30.

Figure 4(b) depicts scatter plots of the ensemble average of the position of the walker at time step $t$, $\langle x(t)\rangle$, versus that of time step $t + D$, $\langle x(t+D)\rangle$, with $D$ being 10,000 considering the five kinds of modulation signals. The blue and cyan marks are based on the experimental results using chaotic modulations (4) Chaos 1 and (5) Chaos 2, respectively, whereas the magenta, red, and green marks correspond to (1) Constant, (2) Periodic, and (3) Quasiperiodic signals. It can be observed that, with chaotically modulated single photons, the walker travels a larger area in the phase diagram of $\left(\langle x(t)\rangle, \langle x(t+D)\rangle\right)$ whereas its diffusing space is rather limited with the other modulations. These observations are consistent with the numerical studies described in Ref. 11.

To summarize, we experimentally examined the concept of dynamically changing probabilistic behavior by combining single photons emitted from a NV center in a nanodiamond and an electro-optical phase modulator. The single photon measurements made by either of the two photodetectors corresponding to horizontally and vertically polarized light directly generated random walkers, whose diffusivities were analyzed using the mean square displacement as well as by examining the trajectories of the walkers in a phase space. Super-diffusion-like behavior was observed on a short time scale by chaotic modulation of single photons. The results agree well with the theoretical and numerical findings reported in Ref. 11. Although this is a scaled-up, proof-of-concept study, it is noteworthy that large diffusivity was experimentally achievable using a rather simple experimental setup, motivating future implementation to nanoscale optical energy transfer, as originally discussed in Ref. 10. A relevant theoretical examination of nano-scale oscillator is the principle using superradiance proposed by Shojiguchi *et al.*[17]. The experimental realization would in either method, however, require breakthroughs in device and material levels; exploiting intrinsic attributes in molecular level, such as photochromism,[18] could be one interesting resource. Meanwhile, chaotically oscillating dynamics have been utilized in recent intelligent functions such as decision making[12], Monte-Carlo computation[19], and artificial data generation through generative adversarial networks[20]; these applications indicate the possibility of directly utilizing random physical processes in nature.

**Acknowledgments**

This work was supported in part by the CREST project (JPMJCR17N2) funded by the Japan Science and Technology Agency, the Core-to-Core Program A. Advanced Research Networks and Grants-in-Aid for Scientific Research (A) (JP17H01277) funded by the Japan Society for the Promotion of Science and Agence Nationale de la Recherche, France, through the TWIN project (Grant No. ANR-14-CE26-0001-01-TWIN) and Placore project (Grant No. ANR-13-BS10-0007-PlaCoRe).

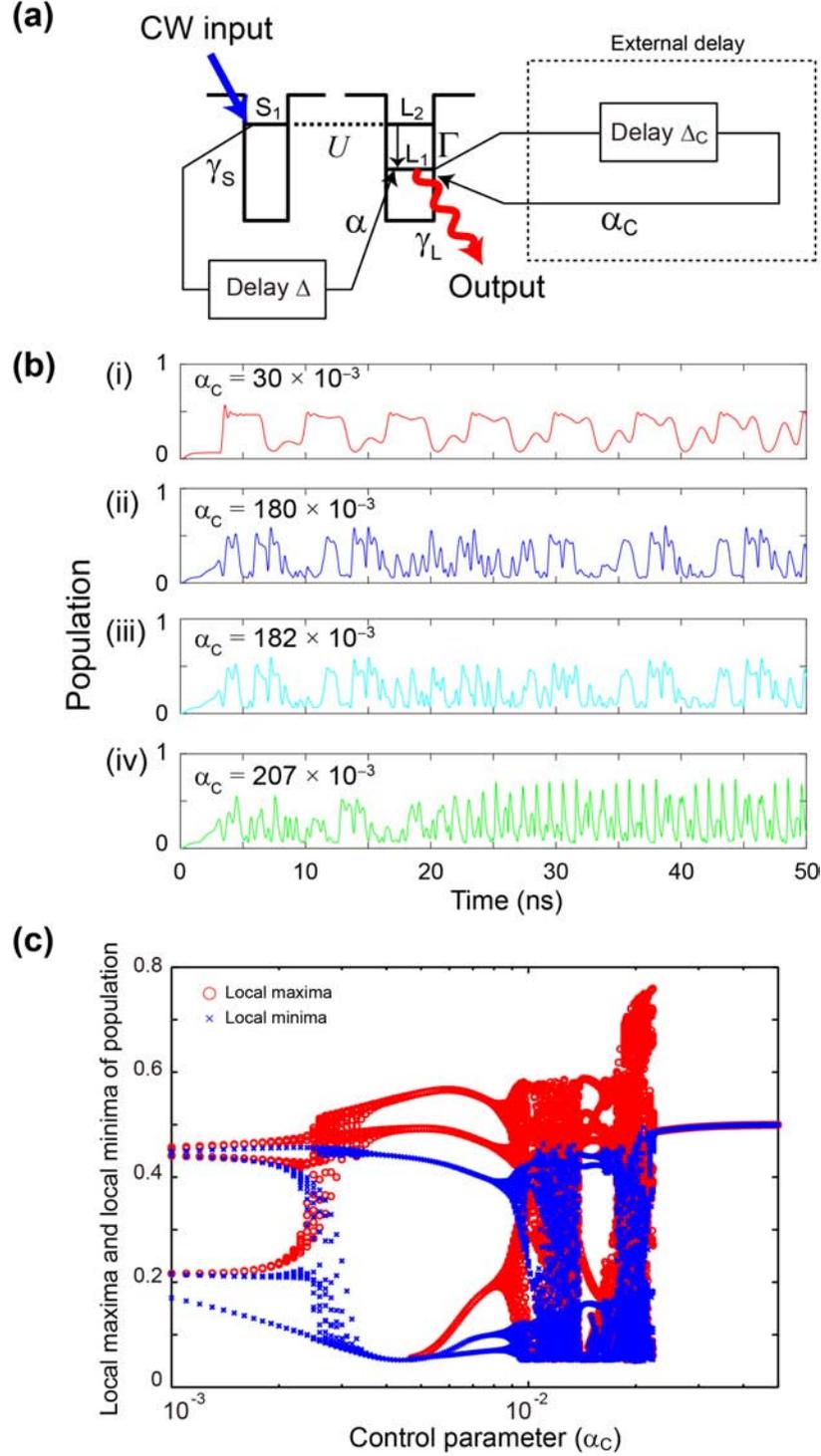

**Fig. 1.** (Color online) Chaos in nano-optical pulser. (a) When optical excitation transfer via near-field interactions between smaller and larger quantum dot involves a time delay Δ, optical pulsation becomes possible.[9] Combined with additional external delay $\Delta_C$, irregular dynamics appears, including chaos.[10] (b) Evolutions of populations from the system: (i) periodic, (ii) quasiperiodic, and (iii) and (iv) chaotic.[10]. (c) With the change in control parameter $\alpha_C$, a clear bifurcation diagram is observed[10]. The images were adapted from Naruse et al., Sci. Rep. 4, 6039 (2014). Copyright 2014 Author(s), licensed under a Creative Commons Attribution 4.0 License.

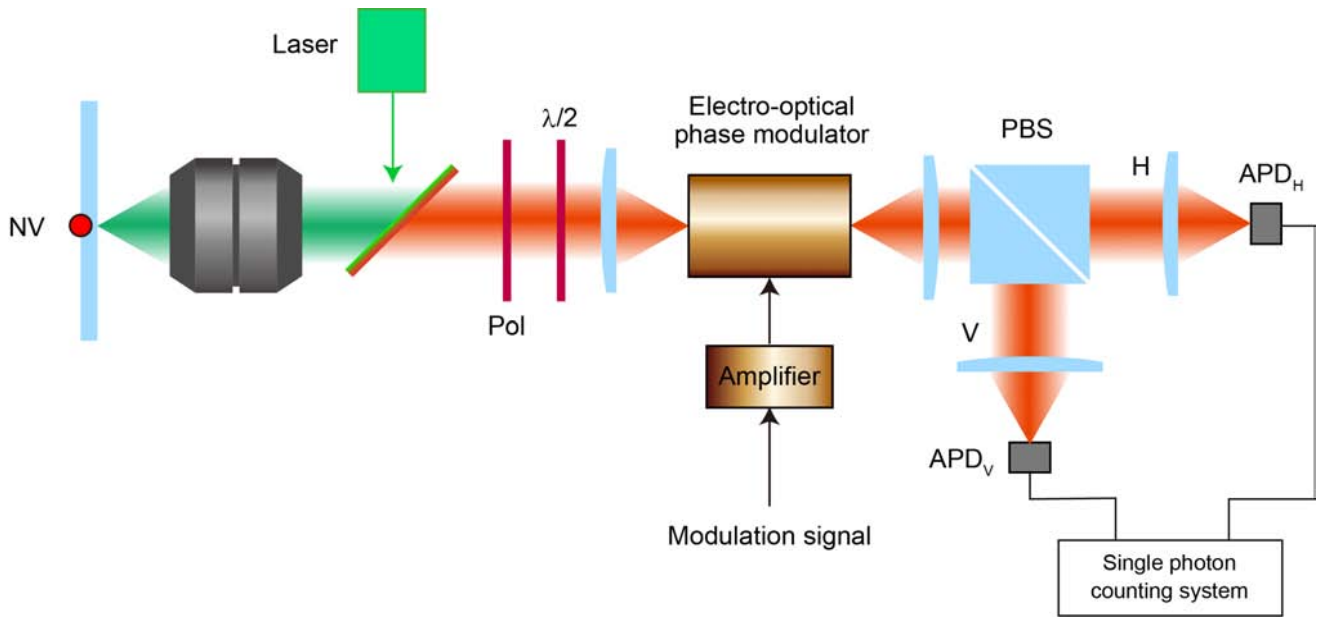

**Fig. 2.** (Color online) Schematic diagram of the experimental setup.

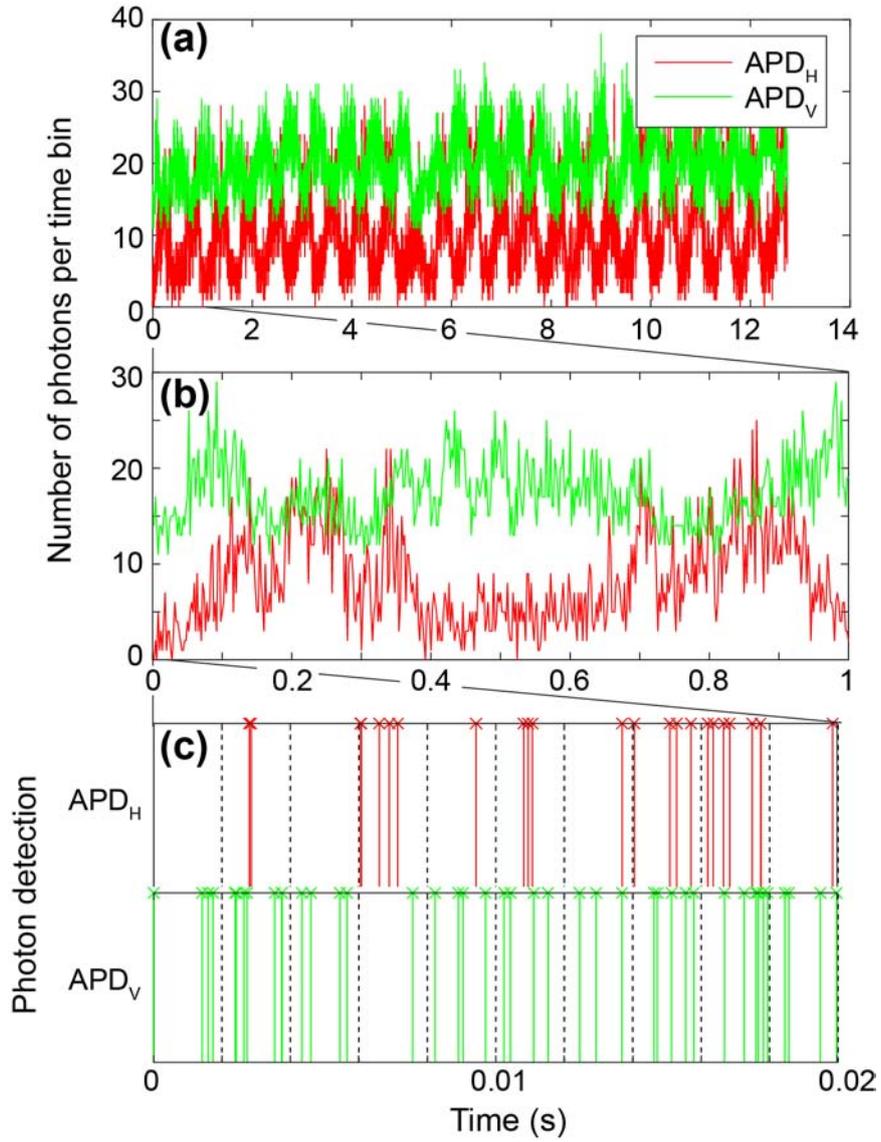

**Fig. 3.** (Color online) (a) Time evolution of single photon detection by $APD_H$ and $APD_V$ when the electro-optical phase modulator was driven by a periodic signal (Fig. 1(b,i)). (b) Magnified view of (a). The vertical axis represents the number of photons per time bin (0.002 s). (c) Single-photon detection events in the initial 0.02 s.

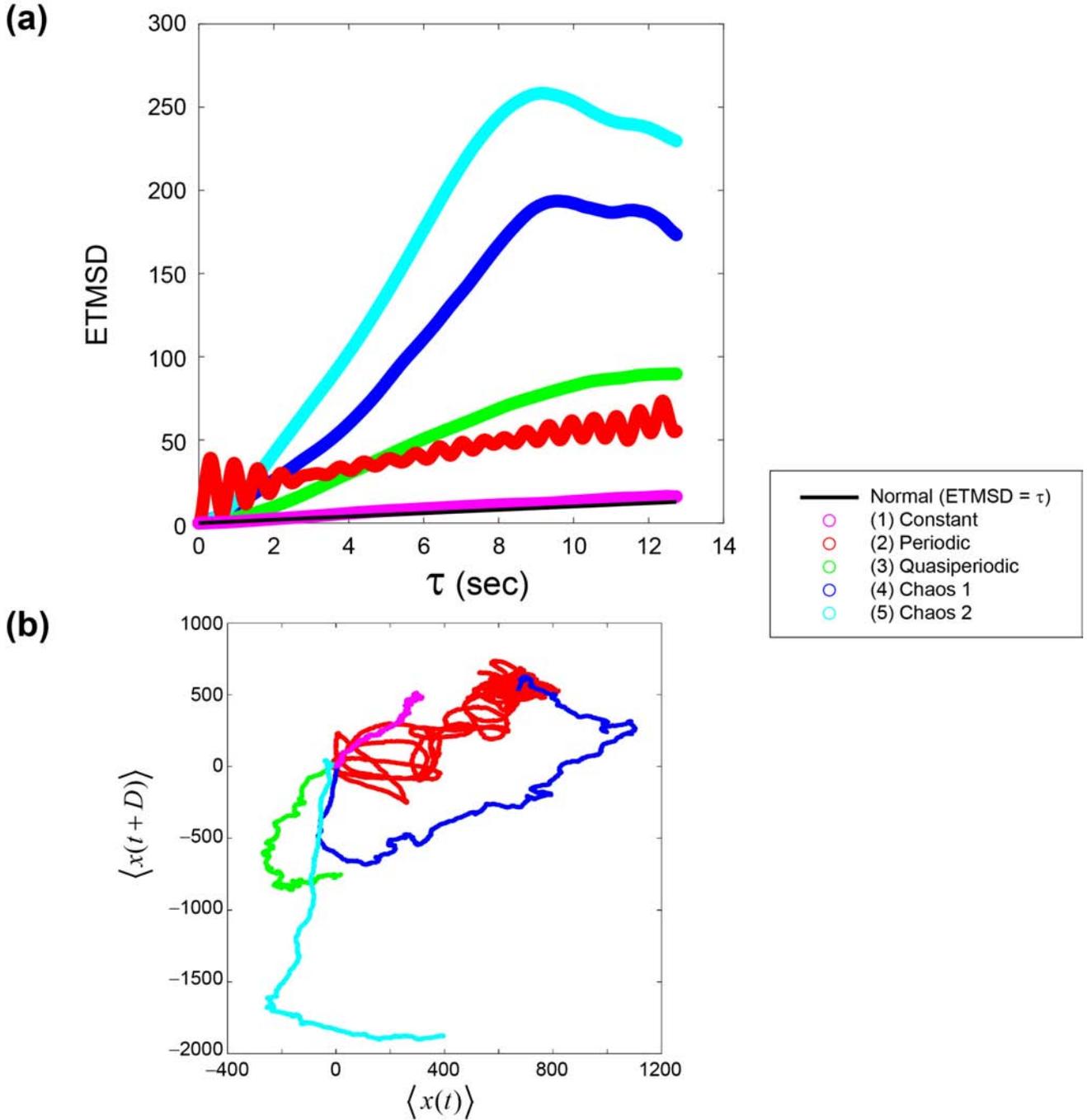

**Fig. 4.** (Color online) (a) ETMSD obtained via the random walker obtained directly from single photon measurements. With chaotic modulation of single photons, a large deviation from the normal scaling is observed. (b) Trajectory of the ensemble average of the position of the walker with the *x*- and *y*-axes representing time *t* and *t* + *D*. With chaotic modulation, the trajectory moves in a large area in the space, which agrees with the numerical study described in Ref. 11.